\documentclass[twocolumn,trackchanges]{aastex61}
\usepackage{natbib}

\begin{document}

\title{Rapid evolution of the gaseous exoplanetary debris around the White Dwarf Star HE 1349--2305}

\author{E. Dennihy}
\affil{Physics and Astronomy Department, University of North Carolina at Chapel Hill, Chapel Hill, NC 27599}

\author{J. C. Clemens}
\affil{Physics and Astronomy Department, University of North Carolina at Chapel Hill, Chapel Hill, NC 27599}

\author{B. H. Dunlap}
\affil{Physics and Astronomy Department, University of North Carolina at Chapel Hill, Chapel Hill, NC 27599}

\author{S. M. Fanale}
\affil{Physics and Astronomy Department, University of North Carolina at Chapel Hill, Chapel Hill, NC 27599}

\author{J. T. Fuchs}
\affil{Department of Physics, Texas Lutheran University, Seguin, TX 78155}

\author{J. J. Hermes}
\affil{Physics and Astronomy Department, University of North Carolina at Chapel Hill, Chapel Hill, NC 27599}
\affil{Hubble Fellow}

\begin{abstract}
Observations of heavy metal pollution in white dwarf stars indicate that metal-rich planetesimals are frequently scattered into star-grazing orbits, tidally disrupted, and accreted onto the white dwarf surface, offering direct insight into the dynamical evolution of post-main-sequence exoplanetary systems. Emission lines from the gaseous debris in the accretion disks of some of these systems show variations on timescales of decades, and have been interpreted as the general relativistic precession of recently formed, elliptical disk. Here we present a comprehensive spectroscopic monitoring campaign of the calcium infrared triplet emission in one system, HE 1349--2305, which shows morphological emission profile variations suggestive of a precessing, asymmetric intensity pattern. The emission profiles are shown to vary on a timescale of one to two years, which is an order of magnitude shorter than what has been observed in other similar systems. We demonstrate that this timescale is likely incompatible with general relativistic precession, and consider alternative explanations for the rapid evolution including the propagation of density waves within the gaseous debris. We conclude with recommendations for follow-up observations, and discuss how the rapid evolution of the gaseous debris in HE 1349--2305 could be leveraged to test theories of exoplanetary debris disk evolution around white dwarf stars.
\end{abstract} 


\section{Introduction}
White dwarf stars have recently emerged as exoplanetary laboratories with the ability to provide detailed elemental abundance ratios via the accretion of tidally disrupted rocky planetesimals \citep{jur14}. Recent results include the measurement of the carbon to oxygen ratio in an unbiased sample of exoplanetary host stars \citep{wil16}, discoveries of water-rich planetesimals \citep{far13,rad15}, and evidence of $^{26}$Al isotopic ratios similar to our solar system \citep{jur13}. 

The dynamical interactions which scatter rocky bodies onto star-grazing orbits that pass within the tidal disruption radius of the white dwarf require both a distant source of rocky material and a large body to facilitate the scattering \citep{deb12,fre14,bon15}. Once the scattered rocky bodies are tidally disrupted, the remnant debris settles into a compact accretion disk which slowly deposits material onto the white dwarf surface \citep{raf12,ver14,ken17a}. The disks mediate the transport of material from the remnant planetary system to the white dwarf, and an understanding of their global evolution can aid studies which translate the observed atmospheric abundances into abundance ratios of the exoplanetary remnants. 

These debris disks were initially identified through the infrared radiation emitted from the dusty, disrupted material \citep{jur03,kil06,von07}. The eventual discovery of double-peaked calcium infrared triplet emission lines \citep{gan06}, a distinctive signature of an orbiting metal-rich gas disk, provided a new avenue to study the debris disks. Despite extensive searches, only seven gaseous emission line systems have since been found and confirmed \citep{gan07,gan08,mel10,duf12,far12,wil14,guo15}. All eight of the gaseous emission systems exhibit significant atmospheric accretion and strong, dusty infrared excesses \citep{far16,den17}. 

The broad spectral features of the emission lines enable more detailed modeling than their dusty infrared excess counterparts, and regular ground-based spectroscopic monitoring campaigns can be used to probe a wide range of evolutionary timescales. To date, variability in the calcium infrared triplet emission profiles at 8498, 8542, and 8662 \AA\ (hereafter \ion{Ca}{2} triplet) has been seen in four of the eight systems, with three systems showing a velocity profile asymmetry shifting from red to blue dominated on timescales of 10-30 years \citep{wil15,man16a,man16b}, and one showing emission lines disappearing completely over the course of 15 years \citep{wil14}. These decadal timescales are challenging to interpret, as they are intermediate to the dynamical timescales of the orbiting exoplanetary debris, which is on the order of hours, and the expected lifetimes of the debris disks, which could be as long as 10$^5$ years \citep{gir12,met12}, though the general relativistic precession of an elliptical disk has been proposed \citep{man16a}. 

Here we present spectroscopic follow-up of the gaseous debris disk orbiting the dusty, metal-polluted, helium-dominated atmosphere white dwarf HE 1349--2305 \citep{gir12,mel12}. Our observations show morphological variations in the \ion{Ca}{2} triplet emission profiles on a timescale of one to two years. Despite many similarities with other well-studied systems, the observed timescale is an order of magnitude shorter, and likely incompatible with the general relativistic precession previously invoked to explain the decadal variations seen in other systems. We consider alternative mechanisms to explain the rapid emission profile evolution, including the propagation of global density waves in the gaseous debris. Finally, we suggest follow-up observations that could support or rule-out the different interpretations we suggest.

\begin{figure*}
\plotone{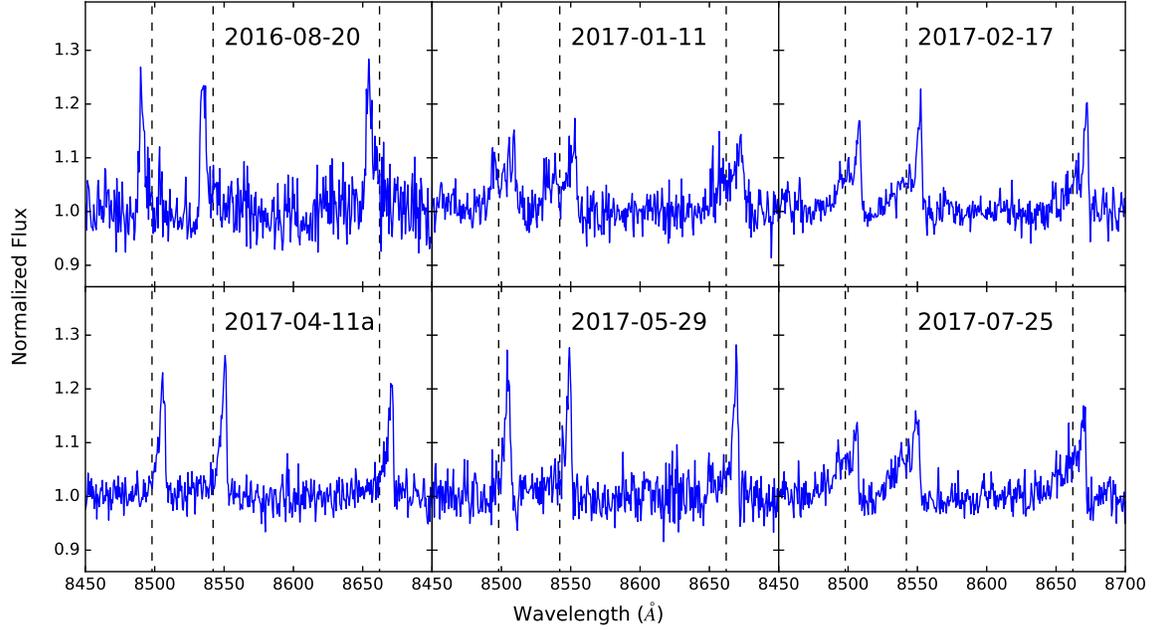}
\caption{Collection of HE 1349--2305 spectra, representative of evolution of the \ion{Ca}{2} triplet emission profiles. The vertical dashed lines denote the rest wavelengths of the \ion{Ca}{2} transitions at 8498, 8542, and 8662\AA. The bulk of the emitting material has undergone a 400 km/s velocity shift in less than nine months.}
\end{figure*}

\begin{deluxetable*}{lcccccccr}
\tablecaption{HE 1349--2305 Observations and \ion{Ca}{2} triplet Emission Profile Measurements}
\tablecolumns{4}
\tablehead{\nocolhead{} &\nocolhead{} & \nocolhead{} & \multicolumn{2}{c}{\ion{Ca}{2} 8498 \AA} & \multicolumn{2}{c}{\ion{Ca}{2} 8542 \AA} & \multicolumn{2}{c}{\ion{Ca}{2} 8662 \AA} \\ \colhead{Date} & \colhead{MJD} & \colhead{Average SNR} & \colhead{EQW} & \colhead{Centroid Vel} & \colhead{EQW} & \colhead{Centroid Vel} & \colhead{EQW} & \colhead{Centroid Vel} \\ 
\nocolhead{} & \colhead{(Days)} & \nocolhead{} & \colhead{(\AA)} &\colhead{(km s$^{-1}$)}& \colhead{(\AA)} &\colhead{(km s$^{-1}$)}& \colhead{(\AA )} &\colhead{(km s$^{-1}$)}}
\startdata
2016-08-20  &  57621.0  &  29  &  1.49 $\pm$ 0.35  &  -215 $\pm$ 29  &  1.60 $\pm$ 0.47  &  -215 $\pm$ 29  &  1.64 $\pm$ 0.34  &  -218 $\pm$ 37  \\ 
2017-01-11  &  57765.3  &  45  &  1.47 $\pm$ 0.32  &  160 $\pm$ 62  &  1.67 $\pm$ 0.24  &  121 $\pm$ 79  &  1.61 $\pm$ 0.28  &  156 $\pm$ 50  \\ 
2017-01-24  &  57778.4  &  43  &  1.42 $\pm$ 0.30  &  16 $\pm$ 54  &  1.89 $\pm$ 0.44  &  101 $\pm$ 48  &  1.42 $\pm$ 0.39  &  94 $\pm$ 74  \\ 
2017-02-08  &  57793.3  &  33  &  1.89 $\pm$ 0.56  &  107 $\pm$ 47  &  1.74 $\pm$ 0.59  &  166 $\pm$ 78  &  1.47 $\pm$ 0.22  &  91 $\pm$ 43  \\ 
2017-02-17  &  57801.3  &  58  &  1.56 $\pm$ 0.16  &  144 $\pm$ 33  &  1.55 $\pm$ 0.30  &  181 $\pm$ 71  &  1.07 $\pm$ 0.41  &  252 $\pm$ 62  \\ 
2017-03-10  &  57822.2  &  57  &  1.26 $\pm$ 0.14  &  257 $\pm$ 18  &  1.43 $\pm$ 0.15  &  253 $\pm$ 28  &  1.50 $\pm$ 0.29  &  202 $\pm$ 54  \\ 
2017-03-13  &  57826.2  &  35  &  1.02 $\pm$ 0.29  &  256 $\pm$ 41  &  1.40 $\pm$ 0.29  &  231 $\pm$ 45  &  1.14 $\pm$ 0.42  &  258 $\pm$ 72  \\ 
2017-04-11a  &  57854.2  &  50  &  1.34 $\pm$ 0.31  &  230 $\pm$ 20  &  1.33 $\pm$ 0.27  &  246 $\pm$ 19  &  1.02 $\pm$ 0.31  &  247 $\pm$ 26  \\ 
2017-04-11b  &  57854.3  &  52  &  1.28 $\pm$ 0.40  &  204 $\pm$ 31  &  0.96 $\pm$ 0.53  &  247 $\pm$ 78  &  1.10 $\pm$ 0.37  &  236 $\pm$ 25  \\ 
2017-04-22  &  57866.2  &  41  &  0.97 $\pm$ 0.23  &  223 $\pm$ 18  &  1.64 $\pm$ 0.48  &  207 $\pm$ 29  &  1.42 $\pm$ 0.28  &  227 $\pm$ 22  \\ 
2017-05-29  &  57903.0  &  44  &  0.87 $\pm$ 0.29  &  206 $\pm$ 20  &  1.34 $\pm$ 0.17  &  197 $\pm$ 16  &  1.36 $\pm$ 0.38  &  195 $\pm$ 24  \\ 
2017-06-08  &  57913.1  &  32  &  1.39 $\pm$ 0.56  &  86 $\pm$ 58  &  1.24 $\pm$ 0.38  &  208 $\pm$ 32  &  1.00 $\pm$ 0.45  &  160 $\pm$ 44  \\ 
2017-07-25  &  57959.0  &  55  &  1.14 $\pm$ 0.28  &  41 $\pm$ 56  &  1.53 $\pm$ 0.50  &  64 $\pm$ 41  &  1.16 $\pm$ 0.27  &  64 $\pm$ 46  \\ 
2017-08-02  &  57968.0  &  33  &  1.07 $\pm$ 0.62  &  93 $\pm$ 134  &  1.62 $\pm$ 0.27  &  79 $\pm$ 54  &  1.51 $\pm$ 0.38  &  143 $\pm$ 68  \\ 
\enddata
\end{deluxetable*}

\section{SOAR/Goodman Observations}
As demonstrated in \citet{man16b}, the gaseous components of the exoplanetary disks often exhibit long-term emission profile variability of that requires regular spectroscopic monitoring to capture. To complement existing observations, we have initiated observing campaigns of the known gaseous debris disk hosting systems with the recently upgraded Goodman Spectrograph \citep{cle04} on the SOAR telescope. The Goodman Spectrograph was designed with an emphasis on ultra-violet and optical throughput, and, as a result, suffered from high-amplitude fringing at wavelengths beyond 750nm. As part of a jointly funded SOAR/NSF project, we redesigned the Goodman Spectrograph to support a second camera outfitted with a deep-depletion fringe-suppressing CCD, enabling the observations presented here.

Our initial observations of HE 1349--2305 are shown in the first panel of Figure 1. The blue-dominated emission peaks represented a significant change from the mildly red dominated peaks presented in \citet{mel12}, and suggested morphological evolution similar to other well studied gaseous debris disks. Since the target was setting, our next observation did not come until January 2017. During the observable season of January 2017 to August 2017, we conducted a comprehensive spectroscopic campaign on HE 1349--2305, collecting data over 14 different epochs with separations ranging from a few hours to a few weeks. The observations are detailed in Table 1, along with several measurements of the \ion{Ca}{2} triplet emission line profiles which are discussed in Section 3. 

For each observation we used the 1200 l mm$^{-1}$ grating and the 1.03\arcsec\ slit for a resolution of approximately 2.25 \AA\ and a wavelength coverage of 7900--9000 \AA. The spectra from each epoch were bias-subtracted, flat-fielded, wavelength-calibrated, and optimally-extracted using a modified version of the Python reduction package described in \citet{fuc17}, before being combined into a final signal-to-noise weighted spectrum. Total integration times ranged from 3600 to 7200 seconds resulting in signal-to-noise ratios in the continuum around 8400 \AA\ between 30 and 50 per pixel. Wavelength calibration was performed using the plethora of night sky emission lines available, as identified by \cite{ost96}, and we have applied heliocentric velocity corrections to each observation.

\section{\ion{Ca}{2} Triplet Emission Profile Measurements} 
Throughout the observing campaign, the \ion{Ca}{2} triplet emission profiles undergo clear morphological changes, starting with a blue-dominated profile in August 2016, transitioning to a nearly symmetric profile in January 2017, rapidly evolving to a red-dominated profile by April 2017, and finally returning to a nearly symmetric profile by July 2017. We have not yet covered a complete cycle of variability, but the additional phase observed by \citet{mel12} suggests the emission profiles could be undergoing periodic evolution. We proceed with the assumption of periodic variability, but note that further observation is needed to confirm this behavior.

To track the variability, we determine the cumulative equivalent width of each profile and measure the wavelength at which it reaches its midpoint, which we term the centroid of the profile. This technique was previously utilized to study the wavelength shifts of polycyclic aromatic hydrocarbons in \cite{slo05,slo07}, and similar techniques have been used to study asymmetries in stellar absorption line profiles \citep{gra05}.

\begin{figure}[ht]
\epsscale{1.15}
\plotone{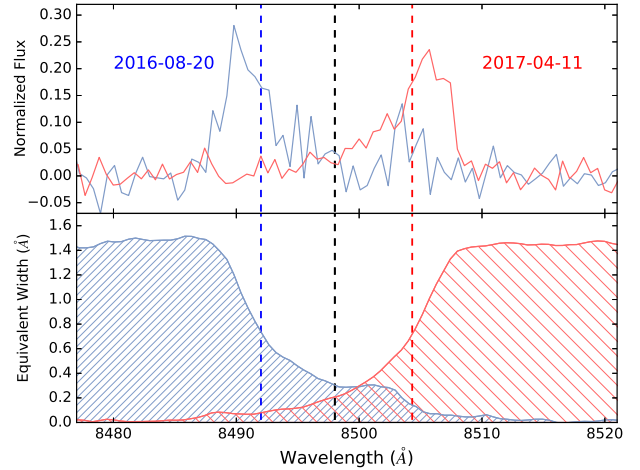}
\caption{Velocity centroid measurements of two epochs with opposite asymmetries. The red and blue dashed vertical lines show the centroid wavelength shifts compared to the rest wavelength of the transition shown in black. The direction of integration is reversed for the 2016 August 20 blue-shifted phase for clarity.}
\end{figure}

Figure 2 demonstrates this centroid measurement for two epochs of opposing phase. In the upper panel we plot the spectra, and in the lower panel we show the cumulative equivalent widths, or integral, of each spectrum as a function of wavelength. Though the of direction integration is reversed for the blue-shifted profile in Figure 2, for our analysis the integration is always performed from blue to red. We translate the wavelength centroids to velocity centroids using the known wavelengths of each transition, providing three independent measurements of centroid velocity for each epoch. 

To derive uncertainties for each velocity centroid measurement, we used a Monte Carlo method of re-sampling the spectra assuming a normal distribution with the measured data as the mean and the rms scatter of the data in the continuum as the standard deviation. We also explored systematic effects of continuum fitting by trialing a range of polynomial orders (linear, quadratic, or third order) and continuum widths (between 25 \AA\  and 30 \AA\  away from the known wavelength of the transition) for normalization of the re-sampled spectra. We include this systematic effect in our final calculation of the values in Table 1, and the measurements and uncertainties represent the mean and range of the velocity centroids found using the different continuum fitting parameters we explored. 

We apply this to the equivalent width measurements and find that they do not vary to within the calculated uncertainties. We note however that this measurement is much more sensitive to continuum fitting than the velocity centroid, and a more careful analysis should be performed before any conclusions on equivalent width variability are drawn. 

We plot our velocity centroid measurements as a function of time in Figure 3. A sinusoidal fit to the velocity centroids with amplitude, phase, period, and mean as free parameters results in a period of 1.4$\pm$0.2 years, which is an order of magnitude shorter than timescales inferred from systems with similar emission profile variations \citep{man16b}. We find no systematic differences in the residuals among the different \ion{Ca}{2} triplet profile measurements. 

We explored the dependence of this period on observation sampling effects by explicitly rejecting the August 2016 measurements and using a Monte Carlo re-sampling of the velocity measurements to explore the distribution of best-fit parameters. Without constraints on the mean or amplitude, the best fit sinusoids prefer shorter periods around 0.8 years. However, these solutions require means and amplitudes around 140  km s$^{-1}$ and 100 km s$^{-1}$, meaning they never result in a phase where the emission profile is blue-shifted, which is confidently ruled out by the three independent August 2016 profiles. If we restrict the mean of the sinusoids to be near zero, we find that the distribution of best fit periods peaks around 1.2 years, with a tail extending to longer periods up to 1.8 years.

Therefore, even in absence of the August 2016 data, the variations support the 1.4 year period interpretation. We note that this interpretation also agrees with our measurement of the velocity centroid for the March 2011 data presented in \citet{mel12}, but the uncertainty of the measured period prohibits an accurate phasing of the two datasets. Finally, we note that we have not yet observed a complete cycle of variations, and it remains to be seen whether the variability is truly periodic. Nonetheless, we assume 1.4 year periodic variations for the discussion below.

\begin{figure}[t]
\epsscale{1.15}
\plotone{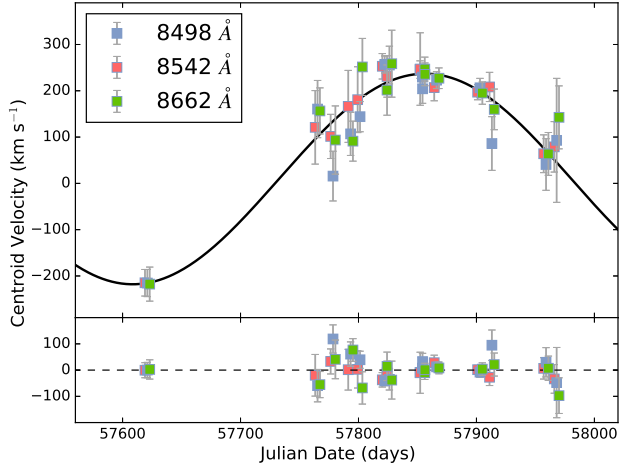}
\caption{The best-fit, 1.4 year period sinusoid (solid line) of the velocity centroid measurements of the \ion{Ca}{2} triplet emission profiles (top) and residuals (bottom).}
\end{figure}

\section{Explanations for the Rapid Variability}

The morphological variability of the asymmetric emission profiles of HE 1349--2305 mimics what has been seen in the gaseous debris disks surrounding WD 1226+110 \citep{man16a}, WD 0845+2257 \citep{wil15}, and WD 1043+0855 \citep{man16b}, all of which have been observed to vary on 10-30 year timescales, though none have observed over a complete cycle of variations. With the confirmation of variability in HE 1349--2305, five of the eight gaseous debris disks around white dwarf stars have undergone significant dynamical changes since their discovery (see Table 2 of \citealt{man16b}). The shorter timescale makes HE 1349--2305 a challenging test case for theories of the variability. 

The leading theory for the asymmetric emission profile variability is the precession of a fixed intensity pattern, which, for the case of WD 1226+110, has been well modeled with both elliptical \citep{man16a} and spiral-shape geometries \citep{har16}. \citet{man16a} further demonstrated that the rate of the gaseous emission profile variability roughly matches the rate expected from the general relativistic precession of mildly eccentric orbits near the midpoint of the debris disk. This interpretation can be similarly applied to the variations observed in the emission profiles of WD 0845+2257 and WD 1043+0855 \citep{man16b}, but  fails in the case of HE 1349--2305. As an alternative, we consider whether the propagation of global density waves could explain both the yearly timescale seen in our observations and the decadal timescales observed in other gaseous emission systems.

\subsection{General Relativistic Precession}
A key component of general relativistic precession is that the timescale is strongly dependent on the orbital radius. In the case of WD 1226+110, the precession timescales range from 1.5 years at the inner edge of the disk to 134 years at the outer edge of the disk, and it is near the midpoint of the disk that the expected precession due to general relativity of 27.8 years matches the observed 24-30 year variations of the emission profiles \citep{man16a}. 

The radial extent of the debris disk in HE 1349--2305 is not as well constrained as in WD 1226+110, but we do have a few reasonable expectations for its inner and outer boundaries. Modeling the gaseous emission lines as optically thick line profiles, \cite{mel12} find the outer radius of the gaseous debris to be near 100 white dwarf radii, and the inner radius, well constrained by the maximum observed velocity of the emission lines, to be between 15 and 20 white dwarf radii depending on inclination. These numbers are broadly consistent with the inner and outer boundaries of the optically thick dust models of \citet{gir12}, which are used to describe the observed infrared excess of HE 1349--2305. These inferred inner and outer radii also match the physical expectations for the boundaries of tidally disrupted debris disks around white dwarfs, where the outer edge corresponds to the tidal disruption radius and the inner edge is set by the dust sublimation radius \citep{far16}.

Following the \cite{man16a} analysis of WD 1226+110 and adopting a white dwarf mass of 0.67 M$_{\odot}$ as determined spectroscopically by \cite{mel12}, in the limit of small eccentricities we find the precessional period due to general relativity at the midpoint of the disk at 60 R$_{\rm WD}$ to be 37.2 years, far too long to explain the 1.4 year variations observed. In order to match the 1.4 year period observed, the emitting gaseous debris would have to be concentrated at 16 R$_{\rm WD}$, which is inconsistent with the extent of the gaseous debris as modeled by \cite{mel12}.

If we relax the assumption of small eccentricity, we can ask at what eccentricity does the precessional period expected from general relativity match the expected midpoint of the debris disk? The eccentricity required for orbits at the midpoint of the disk to precess at 1.4 years is \emph{e} = 0.998, which would carry material from the inner edge of the debris disk at 20 R$_{\rm WD}$ all the way out to $\approx$ 1AU. Such an extremely eccentric orbit is not unexpected during the initial tidal disruption of the larger bodies that supply the debris disk \citep{ver14}, and there is an expectation that if the gaseous debris is constantly being replenished by collisions within the solid debris disk its evolution will mirror that of the solid debris \citep{man16a}, but it is unknown how long such an eccentricity could be sustained in the solid debris under the effects of Poynting-Roberston drag and collisional cascades \citep{ver14,ken17a}.

We have also additional independent geometric constraints on the eccentricity from the observed infrared excess if we assume the dust and gas are co-orbital. This assumption is justified as differences in the eccentricity of the orbits of the dust and gas can lead to runaway accretion scenarios which rapidly deplete the debris disk \citep{met12}. Using the elliptical dust ring models defined in \cite{den16}, we modeled the infrared excess with extremely eccentric rings (\emph{e} $>$ 0.9) and find that they require a near face-on configuration, which contradicts the expected inclination of the gaseous debris as modeled by \cite{mel12}. Therefore, it is not feasible for the dust and gas to be both co-planar and highly eccentric. 

In summary, in order for general relativistic precession to explain the observed timescale of emission profile variations the emitting gaseous material would need to be concentrated in a narrow ring between 16 and 20 R$_{\rm WD}$. Such a compact gas distribution would contradict the previous modeling efforts of the gaseous debris by \citep{mel12}, who find that it should extend out to $\approx$ 100 R$_{\rm WD}$. Given this disagreement, we find it unlikely for general relativistic precession to explain the 1.4 year variations observed in HE 1349--2305, but we cannot yet rule it out.

If, for example, the emission profiles can be modeled with a narrow, inner ring of gas, or it can be demonstrated that the general relativistic precession timescales of the inner edge of the debris disk dominate the evolution of the emission profiles, general relativistic precession would remain a viable possibility. Recent modeling efforts of the evolving absorption profiles of the circumstellar gas around WD 1145+017 have demonstrated that general relativistic precession can readily explain evolutionary timescales as short at 5.3 years if the correct geometry is applied \citep{cau17}. Finally, it is worth noting that if the emission profiles are found to be well modeled by a narrow, inner ring of gas, it could mean that HE134--2305 is in a unique state of evolution, as the gaseous debris in other white dwarf debris disks has been found to extend much further and completely overlap with the dusty debris \citep{mel10}. 

\subsection{Global Density Waves}
As an alternative to general relativistic precession, we propose the evolution of the emission line profiles could be governed by the propagation of global density waves within the gaseous debris disk, and dominate the emission of the gaseous debris. By global density waves, we mean any non-axisymmetric density perturbations that propagate through the gaseous disk, such as spiral density patterns observed in the direct imaging of protoplanetary disks (e.g. \citealt{per16}), or the precessing, elliptical density distributions used to explain emission profile variability in B[e] stars \citep{oka16}. Pattern speeds of these density perturbations can depend on viscosity prescriptions, mechanisms responsible for excitation, and geometric constraints, and thus offer greater flexibility than general relativistic precession.

Spiral density waves observed in protoplanetary disks are often interpreted as evidence of disk-planet interactions with unseen, perturbing bodies \citep{fun15}. For spiral density waves excited by external perturbers on circular orbits, the pattern speed of the wave is expected to match the orbital frequency of the perturber, meaning the variations we have observed would correspond to a perturbing body at a distance of 1-2 AU. Given that the debris disk is completely contained within 1 R$_{\odot}$, a perturbing body at this distance should have little to no effect on the disk. It is worth noting however that spiral density waves can also be excited by recent flyby events, or perturbing bodies on highly eccentric orbits \citep{don15}, though hydrodynamical simulations would be needed to place constraints on these interactions, which is beyond the scope of this paper.

Simulations of turbulence in generalized accretion disks have also shown that disk instabilities such as the magneto-rotational instability readily excite spiral density waves \citep{hei09a,hei09b}. The propagation speed of these waves is coupled to the local sound speed of the gas, though there is some dependence on whether the density waves are treated with non-linear effects \citep{hei12}.

Though their excitation mechanism is largely unknown \citep{oka16}, the one-armed density waves proposed to explain the emission profile variability of the disks observed in B[e] stars offer a more directly analogy to our observations, and similarities between the asymmetric emission profiles observed in B[e] stars and the gaseous debris disks around white dwarf stars were first noted by \cite{gan06}. The timescales observed for emission profile variations in B[e] stars range in period from years to decades, which, similarly to the debris disks around white dwarf stars, is several orders of magnitude longer than the orbital timescales of particles within the disks \citep{oka91}. There is observational evidence that longer emission profile variations correspond to larger disks \citep{rei05}, though the data suggest the radial dependence of this effect is less dramatic than general relativistic precession.

The challenge to this interpretation is likely to be whether density waves can survive long enough to explain the observed variations, particularly given the interactions between the gaseous and dusty debris, which are believed to be spatially coincident \citep{mel10}. If the gas and dust are strongly coupled, density enhancements in the gaseous debris could be rapidly suppressed. Attempts to model this interaction in the environments around white dwarf stars have shown that strong coupling between the gas and dust leads to runaway accretion events which rapidly deplete the debris disk \citep{met12}. This runaway accretion occurs on timescales shorter than the baseline over which some gaseous debris disks have been observed to vary \citep{man16a}, suggesting the coupling in these disks is relatively weak. Continued driving of the density enhancements could also allow them to survive in the presence of damping. 

\section{Recommendations for Follow-up}
High-resolution spectroscopic observations across a range of emission profile asymmetry phases could provide direct constraints on the evolution of material at the inner edge of the disk by tracing the maximum red-shifted and blue-shifted velocities. Correlations of maximum red-shifted and blue-shifted velocities with asymmetry phase have been observed in several B[e] stars \citep{oka16}, and might already be evident in the collected spectra of WD 1226+110 \citep{man16a}. 

The short timescale of the variations observed in HE 1349--2305 also make it an excellent candidate to search for cycle-to-cycle changes that could help confirm or rule out the various interpretations for the profile variability. For example, cycle lengths of the one-armed density waves observed in B[e] disks have been observed to vary sporadically \citep{riv13}. There is also an expectation that global density enhancements could dissipate on viscous timescales without additional driving \citep{ogi01}, which are on the order of years for reasonable assumptions of white dwarf debris disks \citep{man16b}. Changes in the amplitude or period of the emission profile variations of HE 1349--2305 over multiple cycles could be evidence of these processes, contradicting the smooth evolution expected under the general relativistic precession of an elliptical disk.

If, on the other hand, stable periodic variability can be confirmed and the emission lines can be adequately modeled with a much narrower gas distribution, general relativistic precession could remain the preferred explanation for the emission profile variations we have seen. The recent discovery of absorption profile variations around WD1145+017, which were well modeled with an eccentric disk precessing due to general relativity \citep{cau17}, suggests that short precession timescales from general relatvistic precession alone are possible with the correct geometric model, though the even shorter timescale observed in HE1349--2305 may prove challenging to reproduce. Several cycles will be needed to confirm the stable periodic variability, but the short timescale of the variations observed in HE 1349--2305 makes these tests possible over the next few years.

\acknowledgments
We thank Greg Sloan for useful discussions on the centroid method for measuring shifts in emission and absorption profiles. We would like to thank the anonymous referee for helpful comments, criticisms, and questions which improved this manuscript. We would also like to acknowledge the support of the SOAR scientific staff, day crew, and telescope operators. Much of this data was collected during the commissioning of the new Red Camera on the Goodman Spectrograph and without their support during engineering windows these observations would not have been possible. E. D. acknowledges support from the Royster Society of Fellows of the UNC-Chapel Hill Graduate School. E. D., J. C. C., and J. T. F. also acknowledge support from the National Science Foundation, under award AST-1413001. Support for this work was also provided by NASA through Hubble Fellowship grant \#HST-HF2-51357.001-A, awarded by the Space Telescope Science Institute, which is operated by the Association of Universities for Research in Astronomy, Incorporated, under NASA contract NAS5-26555. This work is based on data obtained from the Southern Astrophysical Research (SOAR) telescope, which is a joint project of the Minist\'erio da Ci\^encia, Tecnologia, e Inova\c{c}\~{a}o (MCTI) da Rep\'ublica Federativa do Brasil, the U.S. National Optical Astronomy Observatory (NOAO), the University of North Carolina at Chapel Hill (UNC), and Michigan State University (MSU). 

\software{Astropy \citep{apy13}, ZZCeti Pipeline \citep{fuc17}}

\end{document}